\documentclass[
aps,
prx,
twocolumn,
superscriptaddress,
nofootinbib,
floatfix,
noeprint,
]{revtex4-2}

\usepackage{amsmath,amssymb}
\usepackage{mathtools}
\usepackage{physics}
\usepackage{graphicx}
\usepackage{adjustbox}
\usepackage{tikz}
\usetikzlibrary{quantikz2}
\usepackage{xfrac}
\usepackage[dvipsnames]{xcolor}
\usepackage{float}
\usepackage{bm}
\usepackage{dsfont}
\usepackage{hyperref}
\usepackage{bibunits}
\usepackage{booktabs,tabularx}
\usepackage{etoolbox}
\newcolumntype{C}{>{\centering\arraybackslash}X}

\defaultbibliography{references}
\defaultbibliographystyle{apsrev4-2}

\usepackage{hyperref}
\hypersetup{
	colorlinks=true,
	linkcolor=teal,
	citecolor=teal,
	filecolor=teal,
	urlcolor=teal
}

\newcommand{\TUM}{\affiliation{Technical University of Munich, TUM School of Natural Sciences, Physics Department, 85748 Garching, Germany}}
\newcommand{\MCQST}{\affiliation{Munich Center for Quantum Science and Technology (MCQST), Schellingstr. 4, 80799 M{\"u}nchen, Germany}}
\newcommand{\Nottingham}{\affiliation{School of Physics and Astronomy, University of Nottingham, Nottingham, NG7 2RD, UK}}
\newcommand{\CQNE}{\affiliation{Centre for the Mathematics and Theoretical Physics of Quantum Non-Equilibrium Systems, University of Nottingham, Nottingham, NG7 2RD, UK}}
\newcommand{\Quantinuum}{\affiliation{Quantinuum, Leopoldstrasse 180, 80804 Munich, Germany}}
\newcommand{\UCL}{\affiliation{London Centre for Nanotechnology, University College London, Gordon St., London WC1H 0AH, United Kingdom}}

\begin{document}

\newcommand{\adam}[1]{{\color{red}[Adam: #1]}} 
\newcommand{\hd}[1]{\textcolor{teal}{[HD] #1}}
\newcommand{\shl}[1]{\textcolor{blue}{[SHL] #1}}
\newcommand{\anna}[1]{{\color{ForestGreen}[Anna: #1]}}

\title{Quantum trajectory simulation of two-dimensional non-equilibrium steady states with a trapped ion quantum processor}

\author{Anna Dalmasso} \Nottingham \CQNE
\author{Arash Jafarizadeh} \Nottingham \CQNE
\author{Julian Boesl} \TUM \MCQST
\author{Jared Jeyaretnam} \Nottingham \CQNE
\author{Sheng-Hsuan Lin} \Quantinuum
\author{Andrew G. Green} \UCL
\author{Frank Pollmann} \TUM \MCQST
\author{Michael Knap} \TUM \MCQST
\author{Juan P. Garrahan} \Nottingham \CQNE
\author{Henrik Dreyer} \Quantinuum
\author{Adam Gammon-Smith} \Nottingham \CQNE

\date{\today}

\begin{abstract}
Digital quantum computers offer a promising route for studying complex many-body systems that are otherwise inaccessible by their classical counterparts. 
Capabilities including mid-circuit measurements and feedback allow for simulating the dynamics of interacting open quantum systems. 
Using the Quantinuum System Model H1 trapped-ion quantum computer, we experimentally realise quantum trajectories for a two-dimensional system of (interacting) particles---hard-core bosons or fermions---undergoing stochastic driving at a source and drain at opposite corners of a square lattice. We study the non-equilibrium steady state with persistent current resulting from the this in/out flow of particles. The particle statistics, presence of interactions, and introduction of a magnetic field produce measurable effects on the steady state. Our findings highlight the rich physics in this corner driven two-dimensional setup and showcases both the power and current limitations of quantum computers as a platform to study it.
\end{abstract}

\maketitle

\begin{bibunit}

Driven quantum systems represent a frontier in understanding non-equilibrium quantum physics, particularly regarding how complex many-body systems relax. While isolated systems typically thermalize and follow the Eigenstate Thermalization Hypothesis~\cite{Deutsch1991, Srednicki1994, Rigol2008}, recent theoretical interest has surged around non-ergodic phenomena such as many-body localization~\cite{Basko2006, Abanin2019}, time-crystalline dynamics~\cite{Wilczek2012, Zhang2017, Zaletel2023}, and the existence of steady states maintained by external driving~\cite{Oono1998, Eckmann1999}. 
Driven or open quantum systems may additionally exhibit decoherence, dissipation, and the emergence of non-equilibrium steady states (NESS) characterized by non-zero currents. Tensor networks and MPS methods have been successfully used to characterise the NESS dynamics of one dimensional spin and fermionic chains \cite{Prosen2009, Mendoza-Arenas2013}, however, there are few analytical and numerical tools available for studying interacting NESS beyond small system sizes, especially in dimensions higher than one.

Digital quantum computers offer promising platforms to overcome these limitations.
Mid-circuit measurements, reset, and classical feedback are particularly powerful capabilities for simulating open quantum systems.
They provide non-unitary operations without the added sampling cost of post-selection, and have been demonstrated in a range of quantum computing platforms.
These operations have for instance been utilized in efficient many-body state preparation schemes~\cite{Piroli2021, Smith2023, Tantivasadakarn2023, bluvstein_quantum_2022, Mi2024_reduced, iqbal_topological_2024, iqbal_non-abelian_2024, evered_probing_2025}, and measurement induced phase transitions~\cite{Li2019, Skinner2019, Noel2022, Koh2023, Chertkov2023}.

In this work, we utilize Quantinuum’s H1-1 trapped-ion quantum computer to experimentally realize quantum trajectories for a two-dimensional system of interacting particles. We implement a corner-driven protocol on a  $4 \times 4$ square lattice, creating a NESS with persistent 
\begin{figure}[H]
    \centering
    \includegraphics[width=1\columnwidth]{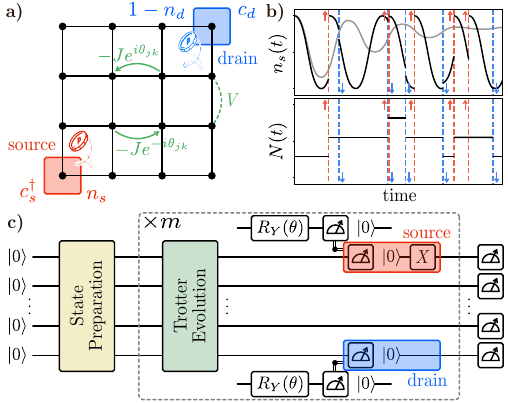}
    \caption{(a) The setup consists of a $4\times4$ square lattice with nearest neighbour hopping (directed arrows) and interactions (dashed line). Stochastic driving acts on the two corner sites labelled ``source" (red) and ``drain" (blue) with probability $p$. (b) Schematic of the quantum trajectory realised by our driving scheme, showing the local density on the source (top) and the total particle number (bottom). The black line is a single trajectory, with source and drain jumps indicated by dashed lines, and grey line indicates the average. (c) The corresponding circuit starts with unitary state preparation followed by unitary evolution using a Trotter decomposition. The non-unitary step uses an ancilla qubit for a ``quantum coin flip" with probability $p$ that classically conditions a measurement and reset on the source and drain. After $m$ timesteps of the unitary and non-unitary drive we perform a full-circuit measurement to extract observables such as local densities and currents.}
    \label{fig:setup_and_circuit}
\end{figure}
current resulting from the injection and removal of hard-core bosons or spinless fermions at opposite corners. 
Particle injection and removal are implemented stochastically using mid-circuit measurements and qubit reset, without the need for post-selection. By tuning particle type, nearest-neighbour interactions, and an external magnetic field, we observe distinct transport regimes, including diffusive flow driven by hydrodynamic modes in the bosonic case and ballistic transport along chiral edge modes in the non-interacting fermionic case. Our results not only highlight the rich physics of this corner-driven two-dimensional setup but also demonstrate both the power and current limitations of near-term quantum computers as a platform to study open system dynamics.

\emph{Setup}---We study a $4 \times 4$ square lattice where we alternate between unitary evolution on the whole system and non-unitary driving at two sites in opposite corners, shown schematically in Fig.~\ref{fig:setup_and_circuit}. The unitary dynamics is defined with respect to the Hamiltonian 
\begin{equation}\label{eq: Hamiltonian}
    H = -J\sum_{\langle j k \rangle} \left( e^{i\theta_{jk}}c^\dagger_j c_k + \text{H.c.} \right) + V\sum_{\langle j k \rangle} n_j n_k,
\end{equation}
where the $c^\dagger$ and $c$ are creation and annihilation operators for either hard-core bosons (HCB) or spinless fermions. We include a magnetic field threading a flux $\phi$ through each plaquette using the Peierls substitution introducing a complex phase into the hopping terms~\cite{Peierls1933}, and a nearest-neighbour density interaction controlled by $V$. We choose $\theta_{jk} = y\phi$ on horizontal bonds, with $y$ the vertical coordinate, and $\theta_{jk}=0$ on vertical bonds. In the unitary part of the drive we evolve with $e^{-i H \Delta t}$ for a time $\Delta t$, which will be implemented in practice using a Trotter decomposition. For the HCB version of the model we directly map the creation / annihilation operators to the qubit raising / lowering operators. For fermions we employ a specific implementation of the Derby-Klassen compact mapping~\cite{Derby2021} (details in Sup. Mat. Section~\ref{sec:fermion mapping}), which requires additional qubits and higher-weight operators, but due to the all-to-all coupling of the H1-1 device results in shallower circuits than the Jordan-Wigner transformation~\cite{Nigmatullin_2025}\footnote{Derby-Klassen and Jordan-Wigner are not the only possible choices for fermion to qubit mapping. We choose Derby-Klassen due to the balance between additional qubits and operator weight.}.

The non-unitary step of the driving stochastically adds particles at the \emph{source} in the bottom left corner, while removing particles from the \emph{drain} in the opposite corner, see Fig.~\ref{fig:setup_and_circuit}.
With probability $p$, we fix the density at the source to be 1, and independently with the same probability $p$ we fix the density at the drain to be 0. We achieve this by using the mid-circuit measurements and reset. For the source, with probability $1-p$ we do nothing; otherwise, we measure that site and store the result, then reset the qubit to the $|1\rangle$ state. This process works identically for both the HCB and fermion mapping. For the drain, we instead leave the reset qubit in the $|0 \rangle$ state. Importantly, this non-unitary step does not require any post-selection. A similar setup was implemented on superconducting qubits in 1D~\cite{Mi2024_reduced, Stenger2025}.

Rather than stochastically generating a set of distinct circuits, we instead utilise two additional ancilla qubits to act as \emph{quantum coins}, one each for the source and drain. We prepare the qubits in the state $R_Y(\theta)|0 \rangle$, where $\theta = \frac{2}{\pi}\arcsin(\sqrt{p})$, leading to a probability $p$ of measuring the $|1\rangle$ state. We then classically condition the measurement and reset processes at the source / drain on measuring the $|1\rangle$ state on the corresponding quantum coin, otherwise they are left untouched. We then reset and prepare these ancilla qubits before each non-unitary phase of the drive. This allows us to define a single circuit where the stochastic trajectory is determined at runtime, as shown in Fig.~\ref{fig:setup_and_circuit}c.

One of the main contributions of our work is that our driving protocol realises quantum trajectories for a dissipative, completely positive, trace-preserving (CPTP) map. More concretely it generates a direct unravelling of the discrete-time map 
\begin{equation}\label{eq:CPTP}
    \rho((n+1)\Delta t) = \sum_k K_k e^{-iH\Delta t} \rho(n\Delta t) e^{iH\Delta t} K_k^\dagger,
\end{equation}
where $\rho$ is the density matrix, and $K_k$ are Kraus operators such that $\sum_k K_k^\dagger K_k = \mathds{1}$. The nine Kraus operators are products $K_k = K^{s}_lK^{d}_m$ of
\begin{equation*}
\begin{aligned}
K^\text{s}_0 &= \sqrt{1-p} \mathds{1},& K^\text{s}_1 &= \sqrt{p} c^\dagger_\text{s}, & K^\text{s}_2 &= \sqrt{p}n_\text{s} \\
K^\text{d}_0 &= \sqrt{1-p} \mathds{1},& K^\text{d}_1 &= \sqrt{p} (1-n_\text{d}), & K^\text{d}_2 &= \sqrt{p}c_\text{d}.
\end{aligned}
\end{equation*}
acting on the source (s) and drain (d). In the limit $\Delta t \rightarrow 0, p/\Delta t \rightarrow \gamma$, this  corresponds to a quantum jump unravelling of the Lindblad master equation
\begin{equation}\label{eq:linblad}
\frac{d \rho}{dt} = -i \left[H, \rho \right] + \sum_i \left( L_i \rho L_i^\dagger -\frac{1}{2}\left\{L^\dagger_i L_i, \rho  \right\} \right),
\end{equation}
where $[X,Y]$ and $\{X, Y\}$ are the commutator and anti-commutator, respectively.
The jump operators are given by
$L_0 = \sqrt{\gamma} n_\text{s}$, 
$L_1 = \sqrt{\gamma} c^\dag_\text{s}$,
$L_2 = \sqrt{\gamma} (1-n_\text{d})$,
$L_3 = \sqrt{\gamma} c_\text{d}$, with the same rate $\gamma$. Due to the form of these jump operators, this model has a non-equilibrium steady state (NESS) with current flowing from source to drain. We set $\gamma=2$, which we determined numerically results in the largest currents (see Sup. Mat. Sec.~\ref{sec:optimal gamma}). In this work we use the Quantinuum H1-1 trapped ion quantum computer to study this NESS. We typically initialise our system in an random product state, corresponding to an infinite temperature mixed state upon averaging, and evolve until the system becomes stationary. For the system sizes we are able to run on H1-1, we perform numerical simulations to find the optimal number of time steps and $\Delta t$ to reach the steady state with minimal circuit depth (see Sup. Mat. Sec.~\ref{sec:optimal parameters}).

\begin{figure}
    \centering
    \includegraphics[width=\columnwidth]{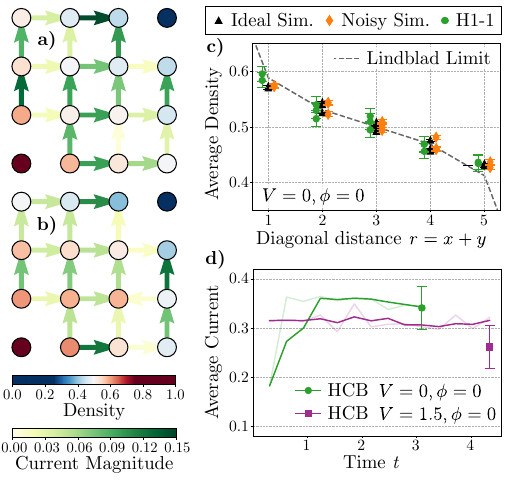}
    \caption{Average local densities and currents for hard-core bosons. (a) Snapshot of the NESS for $V=0$ and $\phi=0$. (b) Snapshot of the NESS for $V=1.5$, $\phi=0$. Both snapshots are measured on the H1-1 device using $\Delta t = 0.31$ and 1280 trajectories, with (a) $m = 10$ and (b) $m=14$ timesteps. (c) The average density as a function of diagonal (Manhattan / taxicab) distance from the source. The density is shown for ideal simulation, noisy emulated simulation (on Quantinuum's H1 emulator) and data from the H1-1 device. Error bars correspond to the standard error of the mean from averaging over trajectories. We additionally include the averaged result close to the Lindblad limit (dashed line, $m = 700$, $dt = 0.01$, $10000$ trajectories). (d) The average current for the two sets of parameters from in/out flow recorded by mid-circuit measurements (faint lines) and their rolling average with a window of 3 time steps (dark lines). The average instantaneous current through the system at the final timestep is also shown (data points).}
    \label{fig:hydro}
\end{figure}

\emph{Results}---We now investigate properties of the NESS under this driving on the Quantinuum H1-1 trapped ion quantum computer. We show results in the NESS for the average density $\langle n_i\rangle$ as well as the local instantaneous current
\begin{equation}\label{eq:current_op}
    \langle \hat{J}_{ij}\rangle = -iJ \langle e^{i\theta_{jk}} c^\dagger_j c_k - e^{-i\theta_{jk}}c_j c^\dagger_k \rangle.
\end{equation}
We run the experiment four times in order to measure the current on four non-overlapping sets of bonds, and a fifth time to measure the local density, and collect statistics for each through many shots. Note that the operator in Eq.~\eqref{eq:current_op} only corresponds to the instantaneous current on bonds not connected to the source and drain, and is approximate for finite $\Delta t$ due to Trotterisation. Results are also compared with ideal and noisy simulations of the same circuit, with the latter run on Quantinuum's H1 emulator, which has an accurate noise model for the device. The ideal simulation data is averaged over 10000 trajectories for both particle statistics, while the noisy simulations used 15000 trajectories for HCB and 6250 for fermions. Each trajectory in the experiment and on the noisy emulator corresponds to a single shot for observables, whereas in the ideal simulation we compute expectation values for each trajectory.

In Fig.~\ref{fig:hydro} we show results for the hard-core boson variant of our driving protocol. Without nearest-neighbour interactions, $V=0$, we observe a density profile varying smoothly from source to drain. This density profile is also shown in Fig.~\ref{fig:hydro}c and is in good agreement with both ideal and noisy simulations. We additionally observe a non-zero current in the NESS, which is dispersed across the system. This behaviour is consistent with the expected hydrodynamic picture for this non-integrable system, with charge carried by hydrodynamic modes~\cite{ChaikinLubensky1995}. This is moreover reflected by the NESS being unaffected by the magnetic field (not shown). 

When including nearest-neighbour interactions, $V=1.5J$, the NESS is shown in Fig.~\ref{fig:hydro}b. Due to the increased time required to relax to the stationary state, we initialize in a randomly selected product state where the probability of each site being filled is proportional to the average density in the NESS found through numerics.
This density interaction breaks the particle-hole symmetry in the bulk, resulting in an average filling $>0.5$. We additionally observe a rearrangement of the density predominantly filling the $3\times3$ square closest to the source, with the remaining two edges having reduced density. This can be explained using an energetic argument: removing a particle from a filled region has an interaction energy cost proportional to the number of nearest-neighbours. Therefore there is a lower energy barrier to removing particles along the edges connected to the drain. In contrast, adding a particle to an empty region has no cost from the interaction term, allowing the source to more efficiently fill the system. 
Numerical simulations of a classical stochastic symmetric simple exclusion process SSEP~\cite{Derrida2007} recovers this density patterns and supports the energetic argument (see Sup. Mat. Sec.~\ref{sec:SSEP}).

The bunching of the density close to the source and low density near the drain blocks the current through the system. In Fig.~\ref{fig:hydro}d, we show the average current as a function of time from the in / out flow of particles as monitored through the midcircuit measurements. This shows that the current is reduced by the presence of the interactions, and agrees with the average instantaneous current across counter-diagonal cuts in the NESS (data points) up to statistical errors. This latter point is non-trivial, since errors in the device could include unwanted particle creation and annihiliation. Note that while the mid-circuit measurements show quick stabilisation of the net current, we performed numerical simulations to motivate our experimental parameters such that the local currents in the bulk are also steady at the end of the simulation (Sup. Mat. Sec.~\ref{sec:optimal parameters}).

\begin{figure}
    \centering
    \includegraphics[width=1\columnwidth]{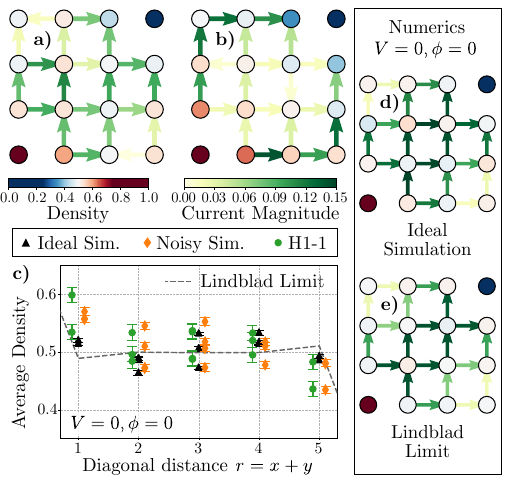}
    \caption{
    (a) Final snapshots for the local density and instantaneous currents measured on the H1-1 device for fermions with $V=0, \phi=0$ ($m=14$ timesteps, $\Delta t = 0.21$), and (b) for $V=0, \phi=\pi/2$ ($m=16$ timesteps, $\Delta t = 0.27$). Both snapshots were averaged over 1480 trajectories. 
    (c) The average density as a function of diagonal (Manhattan / taxicab) distance from the source using ideal simulation, noisy emulated simulation (on Quantinuum's H1 emulator), data from the H1-1 device and the Lindblad limit (dashed line). 
    (d) Shows the snapshot from the ideal simulation for fermions $V=0, \phi=0$, and (e) the corresponding steady state close to the Lindblad limit ($m = 1000$, $dt = 0.01$, $10000$ trajectories).
    } 
    \label{fig:integrable}
\end{figure}

Next, let us consider data from the fermionic variant of the model, shown in Fig.~\ref{fig:integrable}, which we implemented using a modified version of the Derby-Klassen compact mapping \cite{Derby2021} (details in Sup. Mat. Section A). Without interactions or magnetic field (Fig.~\ref{fig:integrable}a) the current is predominantly along the diagonal between the source and drain, with negligible current through the off-diagonal corner sites. Numerics with experimental parameters, as well as for small $\Delta t$ show an even stronger focussing of the current along the diagonal (Fig.~\ref{fig:integrable}d-e). Furthermore, Fig.~\ref{fig:integrable}c demonstrates the flat density profile in the bulk, which is in stark contrast to the gradient observed for HCB. This effect has been studied in one-dimension and has been termed hyperuniformity~\cite{Prosen2011,Torquato2016,Carollo2017,Torquato2018}. These differences in behaviour can be understood through the free-fermion integrability of the Hamiltonian evolution in the bulk. Despite the Lindbladian not being free-fermion solvable, the current is carried by ballistic bulk modes, with strongest occupation of the diagonal mode commensurate with both the source and drain. 

To further emphasize the effects of this ballistic transport, we add a magnetic field with $\phi=\pi/2$ threading each plaquette. We observe dramatic change in the NESS, with currents flowing almost entirely along the boundary of the system, shown in Fig.~\ref{fig:integrable}b for $V=0$. In the absence of the non-unitary step in our drive, our system corresponds to the Harper-Hofstadter model~\cite{Hofstadter1976} with chiral modes in the spectral gap: one chirality at low energy, and the counter propagating mode at high energy. When adding and removing particles we symmetrically occupy both these two modes and have a net achiral current along the edges. In contrast, the NESS for HCB is unchanged by the presence of the magnetic field (shown numerically in Sup. Mat. Sec.~\ref{sec:HCB field}) due to the current in the NESS being carried by collective hydrodynamic modes, which are unaffected by the magnetic field. We stress that while the free-fermion nature of the bulk makes numerical simulations easier, this fine-tuned case is actually more sensitive to errors on the quantum computer, since any unintended scattering would lead to diffusive and not ballistic charge transport~\cite{chertkov2026robustness}. 

\begin{figure}[t]
    \centering
    \includegraphics[width=1\columnwidth]{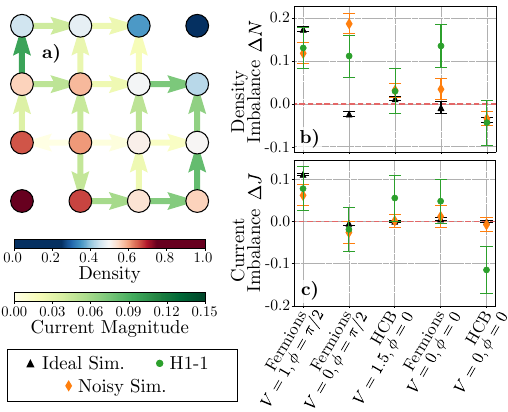}
    \caption{
    (a) Final snapshots for the local density and instantaneous currents measured on the H1-1 device for fermions $V=1.0, \phi=\pi/2$ ($m=18$ timesteps, $\Delta t = 0.29$, 1480 trajectories). (b) Shows the total density imbalance above and below the diagonal between source and drain for the 5 cases we considered, and (c) shows the corresponding current imbalance. Results are shown from ideal simulation (black triangles), noisy simulation (orange diamonds) and data from the H1-1 device (green circles) and the red dashed line indicates zero imbalance. 
    }
    \label{fig:chiral}
\end{figure}

In light of understanding the NESS in terms of either diffusive or ballistic charge transport, we would generically expect the introduction of nearest-neighbour interactions to result in a hydrodynamic description in the thermodynamic limit. Nonetheless, there will be a length scale for the scattering due to interactions, and so for finite system sizes there will remain interesting qualitative differences between HCB and fermions. In Fig.~\ref{fig:chiral}a we include both interactions $V=1$ and magnetic field $\phi=\pi/2$ for fermions, which break the chiral symmetry in the NESS visible in both the density and the current, an effect that is not observed for HCB (see Sup. Mat. Sec.~\ref{sec:HCB field}). First, the density is no longer mirror symmetric with respect to the diagonal between source and drain: we have a higher density in the lower-right corner. Additionally, there is some indication of an anti-clockwise bias to the edge currents. However, both the densities and the currents fluctuate more strongly than when $V=0$, in part due to the additional circuit depth required to reach the NESS (18 timesteps).

To better quantify the chirality when including both magnetic field and interactions, we introduce the density and current imbalances
\begin{alignat}{3}
    \Delta N &=\ \sum_{\mathclap{j \in\,\text{below}}} \langle n_{j} \rangle
    &\;-&& &\;\sum_{\mathclap{j' \in\,\text{above}}} \langle n_{j'} \rangle\,,\\
    \Delta J &=\ \sum_{\mathclap{i, k \in \substack{\text{ edges} \\ \text{ below}}}} \langle \hat{J}_{ik} \rangle
    &\;-&& &\ \sum_{\mathclap{i',k' \in \substack{\text{ edges} \\ \text{ above}}}} \langle \hat{J}_{i'k'} \rangle\,,
\end{alignat}
which compute the difference in density / current above and below the diagonal between the source and drain, where currents are directed from source to drain. The data, shown in Fig.~\ref{fig:chiral}b-c, reveal much about the physics in our systems as well as errors in the H1-1 device. The ideal simulations nicely match our expectations and show zero imbalance in both density and current for all cases except when chiral symmetry is broken, $V=1.0, \phi=\pi/2$, where we see the anti-clockwise bias of the density and current. While the density imbalance is correctly captured on the device in this case, we also measure spurious positive imbalances for fermions $V=0=\phi$ and $V=0, \phi = \pi/2$. Similarly for the current imbalance, while the experimental data correctly identifies the chirality for $V=1, \phi=\pi/2$, it incorrectly has negative value for HCB with $V=0$, $\phi=0$.

Statistical error may play a large role in the discrepancy between ideal simulations and experimental results. Results from the noisy emulator---which simulates an accurate noise model for the device---have improved agreement with the ideal simulation due to the increased number of trajectories (15000 vs. 1280 for HCB, and 6250 vs. 1480 for fermions). Unfortunately, improving the statistics for the experimental data is limited by the speed of gate and shuttling operations on H1-1. Nonetheless, there is a statistically significant outlier for fermions $V=0, \phi=0$. Since the experimental data and noisy simulation are consistent for this case and both disagree with the ideal simulation, it suggests there is some additional bias on the device. One such possibility---that unfortunately we are unable to check---is asymmetric idling time for qubits when the circuits are compiled to native operations.

\emph{Outlook}---In this paper we have utilised the capability of Quantinuum's H1-1 trapped ion quantum computer to study non-equilibrium open quantum dynamics in two dimensions. We devised a driving protocol that implements quantum trajectories for a dissipative CPTP map. By additionally employing a compact fermion-to-qubit mapping, we were able to efficiently simulate fermionic dynamics as well as hard-core bosons. In our simple model we were able to reveal the rich properties of the NESS and its persistent current resulting from hydrodynamics, integrability, interactions and a magnetic field. Our work has demonstrated how mid-circuit measurements, reset and feedback allows us to simulate quantum trajectories and opens up a flexible route for their study on quantum devices. 

One of the most interesting aspects of our results is how the free-fermion integrability of the bulk Hamiltonian results in non-hydrodynamic transport in the NESS, despite the fact that the jump operators break the integrability of the Lindbladian. In particular, we observed the current flowing ballistically along the dominant modes connecting the source and the drain. Furthermore, tuning the magnetic field allowed us to change the path of the current to flow along the boundary. Understanding the symmetries and spectral properties of the Lindblad operator in Eq.~\eqref{eq:linblad} would be a particularly interesting direction for future work.

We have focussed on the stationary state of Eq.~\eqref{eq:linblad}, but we expect the dynamics in this system to be at least as rich. 
Due to practical limitations on the device, we have been unable to properly analyze the statistics of trajectories that we have recorded through our mid-circuit measurement. We would be particularly interested to apply a large deviations approach to study the rare fluctuations in this dynamics~\cite{Garrahan2010}. We also noted that the NESS for HCB was unaffected by the magnetic field, but it is unclear what effect it might have on the dynamics of trajectories.

\vspace{5pt}
\emph{Acknowledgements}---A.D., A.J., and A.G-S. were funded by UK Research and Innovation (UKRI) under the UK government’s Horizon Europe funding guarantee [grant number EP/Y036069/1]. J.J., J.P.G., and A.G-S. acknowledge support from the Leverhulme Trust through Research Project Grant [RPG-2024-112]. J.P.G. acknowledges support from EPSRC [Grant No. EP/V031201/1]. J.B., M.K., F.P. acknowledge support from the Deutsche Forschungsgemeinschaft (DFG, German Research Foundation) under Germany’s Excellence Strategy--EXC--2111--390814868, TRR 360 – 492547816, the European Union (grant agreement No. 101169765), as well as the Munich Quantum Valley (MQV), which is supported by the Bavarian state government with funds from the Hightech Agenda Bayern Plus.
Experimental data reported in this work was produced by Quantinuum H1-1 quantum computer, powered by Honeywell, between 21/07/2025 and 21/10/2025. Access to the device was provided through the Quantinuum promotional access program. We are extremely grateful to the Quantinuum technical and support team for the support they provided during this project.

\vspace{5pt}
\emph{Data Availability}---Code and data are available on Zenodo~\cite{data_code}

\vspace{5pt}
\emph{Competing Interests}---H.D. is a shareholder of Quantinuum.

\putbib
\end{bibunit}

\bibliography{references}

\onecolumngrid
\clearpage
\twocolumngrid

\onecolumngrid 

\begin{center}
  \textbf{\large Supplemental Material for ``Quantum trajectory simulation of two-dimensional non-equilibrium steady states with a trapped ion quantum processor''}\\[.2cm]
\end{center}

\vspace{1cm} 

\twocolumngrid

\begin{bibunit}

\appendix
\renewcommand{\appendixname}{}

\setcounter{figure}{0}
\setcounter{table}{0}
\setcounter{page}{1}
\renewcommand{\thefigure}{S\arabic{figure}}
\renewcommand{\thepage}{\roman{page}}
\renewcommand{\thetable}{S\arabic{table}}

\section{Fermionic Mapping}\label{sec:fermion mapping}
\begin{figure}[t]
    \centering
    \includegraphics[width=0.8\columnwidth]{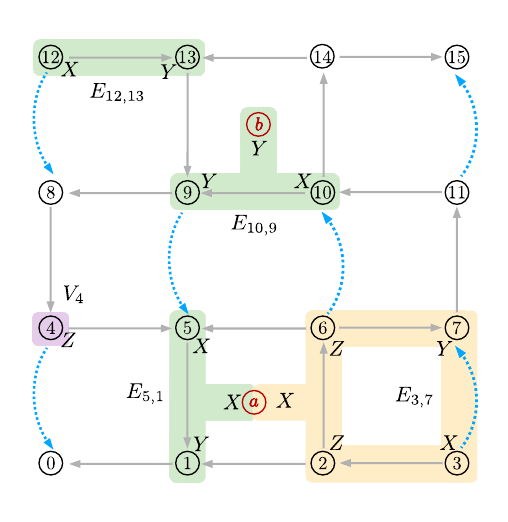}
    \caption{Schematic representation of the hybrid fermionic mapping used in this work. The black vertices correspond to the physical qubits numbered from ``0" (source) to ``15" (drain), the two red circles are ancillary qubits labelled ``a" and ``b". The arrows on the bonds indicate directionality of the local edge operators $\tilde{E}_{i,j}$. The dotted blue arrows correspond to positive Jordan-Wigner strings created from products of neighbouring $\tilde{E}_{i,j}s$.}
    \label{fig:hybrid_mapping_detailed}
\end{figure}
Following the mapping presented in references \cite{Derby2021, Nigmatullin_2025}, we introduce the following edge and vertex operators expressed in terms of Majorana operators $\gamma_j = c_j^\dagger + c_j$ and $\bar{\gamma_j} = i(c_j^\dagger - c_j$):
\begin{align}
    E_{i,j} &= -i \gamma_i\gamma_j\\
    V_{i} &= -i\gamma_i\bar{\gamma_{i}} 
\end{align}
and the corresponding local mapped operators: 
\begin{align}
    \tilde{E}_{i,j} &=
        \begin{cases}
            X_{i}Y_{j}X_{a} & (i, j) \text{ vertical}\\
            X_{i}Y_{j}Y_{a} & (i, j) \text{ horizontal}
        \end{cases}\\
    \tilde{V}_{i} &= Z_{i} 
\end{align}
where $\tilde{E}_{i,j} = -\tilde{E}_{j,i}$. For every oriented edge $(i,j)$, the arrow always points from site $i$ to site $j$ and the subscript $a$ denotes the ancilla qubit directly adjacent to the bond $(i , j)$; in the absence of a neighbouring ancilla, the last Pauli operator $P_a$ is omitted.
These operators are represented by gray arrows in Fig.~\ref{fig:hybrid_mapping_detailed}. 
Hopping between sites that are not connected by gray arrows can be done by applying a product of neighbouring edge operators following the direction of the arrows, with an additional minus sign.
For example, the operator acting on the bond (4, 0) can be derived as follows:
\begin{align}
    \tilde{E}_{4,0} &= - \tilde{E}_{4,5}\tilde{E}_{5,1}\tilde{E}_{1,0}\\
                    &= X_{4}Z_{5}Z_{1}X_{a}Y_{0} 
\end{align}
These longer range edge operators are represented by blue dotted lines in the same figure.
It is easy to show that the mapped operators obey the correct anti-commutation relations:
\begin{align}
    \{ \tilde{E}_{i,j}, \tilde{E}_{j,k} \} =0, && \{\tilde{E}_{i,j}, \tilde{V}_{j} \} = 0,
\end{align}
when they share a vertex, and for all $i \neq j \neq m \neq n$:
\begin{align}
    [\tilde{E}_{i,j}, \tilde{E}_{m,n}] =0, && [\tilde{E}_{i,j}, \tilde{V}_{m} ] = 0 && [\tilde{V}_{i}, \tilde{V}_{j}] = 0 .
\end{align}
To define the physical subspace of this enlarged qubit Hilbert space, we need to obey one additional constraint, that the product of loop operators around all plaquettes equals the identity, namely,
\begin{equation}\label{eq:constraint}
    \prod_{j \in \text{loop}} i \tilde{E}_{j,j+1} = \mathds{1}.
\end{equation}
This constraint is obeyed on almost all plaquettes. The product around the central plaquette is however non trivial and equal to $\mathbb{J}$:
\begin{equation}\label{eq:stabilizer}
    \prod_{j \in [4, 11]} i \tilde{E}_{j,j+1} = \left( \prod_{j \in [4, 11]} Z_{j}\right) Y_{a}Y_{b} = \mathbb{J}.
\end{equation}
This operator corresponds to the stabilizer of the encoding and we identify the physical subspace with its unique +1 eigenspace.

\section{Circuits}
\subsection{Unitary two-qubit gate}
After Trotter decomposition the unitary evolution operator is written in terms of a product of local single and two-qubit operators. In the absence of external magnetic field, the smallest unitary operator expressed in terms of Pauli matrices has the form:
\begin{equation}\label{eq:unitary_operator}
    U_{ij} = e^{i(\frac{J}{2}X_{i}X_{j} + \frac{J}{2}Y_{i}Y_{j} - \frac{V}{4}Z_{i}Z_{j})\Delta t} \ e^{i\frac{V}{4}Z_{i}\Delta t} \ e^{i\frac{V}{4}Z_{j}\Delta t}
\end{equation}
where $i$ and $j$ are two neighbouring sites on the square lattice. Before expressing this operator in terms of a quantum circuit, we define the rotation gates as follows:
\begin{align*}
    R_{Z}(\theta) &= e^{-\frac{i}{2}\pi \theta Z} = 
        \begin{pmatrix}
            e^{-\frac{i}{2}\pi \theta} & 0 \\
            0 & e^{\frac{i}{2}\pi \theta} 
        \end{pmatrix} \\
    R_{X}(\theta) &= e^{-\frac{i}{2}\pi \theta X} =
        \begin{pmatrix}
            \cos{(\frac{\pi \theta}{2})} & -i\sin{(\frac{\pi \theta}{2})} \\
            -i\sin{(\frac{\pi \theta}{2})} & \cos{({\frac{\pi \theta}{2})}} 
        \end{pmatrix} \\
    R_{Y}(\theta) &= e^{-\frac{i}{2}\pi \theta Y} =
        \begin{pmatrix}
            \cos{(\frac{\pi \theta}{2})} & -\sin{(\frac{\pi \theta}{2})} \\
            \sin{(\frac{\pi \theta}{2})} & \cos{(\frac{\pi \theta}{2})} 
        \end{pmatrix}
\end{align*}
where we include a factor of $\pi$ in the definition. Now we can write the two single qubit operators in Eq.~\ref{eq:unitary_operator} as simple $R_{Z}(-\frac{V\Delta t}{2\pi})$ rotations on each qubit. For the two-qubit operator we use the $TK2(\alpha, \beta, \gamma)$ as a native gate. 

To include the magnetic field phase, we need to apply additional single qubit rotations $R_{Z}(\pm \frac{\phi}{2})$ on sites $i$ and $j$ before and after the $TK2$ gate.
Now the smallest two-qubit circuit representing the Hamiltonian from Eq.~\eqref{eq: Hamiltonian} can be written as:
\\
\\
\begin{adjustbox}{width=\columnwidth}
\tikzset{
operator/.append style={minimum size=11mm},
gg label/.append style={font=\large}
}
\begin{quantikz}[wire types={q,q}]
&   \gate[2]{\large{U_{ij}}}   &\midstick[2,brackets=none]{=}&   \gate{R_Z(-\phi/2)} & \gate[2]{\begin{array}{c} \text{\large{TK2}} \\ \text{($\alpha, \beta, \gamma$)} \end{array}} & \gate{R_Z(\phi/2)}    &   \gate{R_Z(-\gamma)}  &
\\
&  && \gate{R_Z(\phi/2)} &  & \gate{R_Z(-\phi/2)}  &   \gate{R_Z(-\gamma)} &
\end{quantikz}
\end{adjustbox}
\\
\\
\\
where $\alpha = \beta = -\frac{J\Delta t}{\pi}$ and $\gamma = \frac{V\Delta t}{2\pi}$.
This unitary gate can be applied on almost all bonds for both fermionic and bosonic circuits. 
Because of our choice of mapping, we need an extra modified unitary gate $\tilde{U}_{ij}$ in the fermionic circuit:
\\
\\
\begin{adjustbox}{width=\columnwidth}
\tikzset{
operator/.append style={minimum size=11mm},
gg label/.append style={font=\large}
}
\begin{quantikz}[wire types={q,q}]
&   \gate[2]{\large{\tilde{U}_{ij}}}   &\midstick[2,brackets=none]{=}&   \gate{R_Z(-\phi/2)} & \gate[2]{\begin{array}{c} \text{\large{TK2}} \\ \text{($-\alpha, -\beta, \gamma$)} \end{array}} & \gate{R_Z(\phi/2)}    &   \gate{R_Z(-\gamma)}  &
\\
&  && \gate{R_Z(\phi/2)} &  & \gate{R_Z(-\phi/2)}  &   \gate{R_Z(-\gamma)} &
\end{quantikz}
\end{adjustbox}
\\
\\
\\
$\tilde{U}_{ij}$ is applied instead of $U_{ij}$ on the bonds $(i,j)=(5,6), (13,14)$ in order to satisfy constraint~\ref{eq:constraint} (see Fig.~\ref{fig:fermionic_trotter}).

\subsection{Bosonic circuit}
\begin{figure}[t]
    \centering
    \includegraphics[width=1\columnwidth]{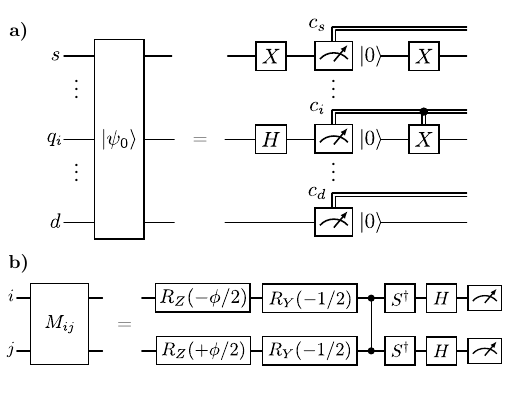}
    \caption{Bosonic state preparation (top) and two-qubit transformation to measure currents (bottom). The qubit labels ``$s$" and ``$d$" refer to source and drain, while the labels ``$i$" and ``$j$" indicate two nearest-neighbour qubits. After initial density measurements we store the outcomes in classical registers ``$c_x$", one for each qubit ``$x$".}
  \label{fig:common_circuits}
\end{figure}

\begin{figure}[t]
    \centering
    \includegraphics[width=\columnwidth]{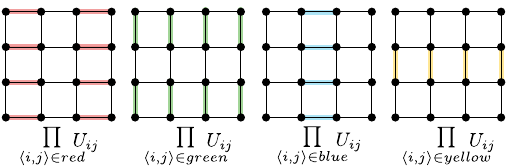}
    \caption{HCB Trotter circuit for one period of evolution. The full Trotter step is made out of four non-overlapping sectors (in different colours) that are applied in order red, green, blue, yellow. The circuit in each sector is a product of local unitary operators $U_{ij}$ applied on different neighbouring qubits $i$ and $j$.}
    \label{fig:HCB_trotter}
\end{figure}

To build the full bosonic circuit, the first step is preparing the initial product state. 
In the absence of nearest-neighbour interactions, we start with random occupations on each site. This can be achieved by applying a single Hadamard gate on every qubit and then fix the occupation by making measurements, as shown in Fig.~\ref{fig:common_circuits}a. To kick-start the driving we also ensure that source and drain are in states  $| 1 \rangle$ and $| 0 \rangle$ respectively.
For $V>0$ we instead prepare the qubits in a density pattern closer to the one expected in the NESS from numerics. This is done by preparing each qubit in the state $R_Y(\alpha_i)$, where $\alpha_i = \frac{2}{\pi}\arcsin{(\sqrt{\langle n_i \rangle})}$ depends on the expectation value of density on site $i$ obtained from numerical simulations.  

We then proceed with trotter evolution by applying the unitary two-qubit gates $U_{ij}$ on neighbouring sites. The bonds are grouped in four non-overlapping sectors and the unitary gates are applied on one sector at a time, as shown in Fig.~\ref{fig:HCB_trotter}.

The full circuit is then built by alternating Trotter evolution and coin flips with potential driving for $m$ timesteps, as shown in Fig.~\ref{fig:setup_and_circuit}c. After we reached the NESS at final time $T = m \Delta t$, we perform full circuit measurements to determine local densities and currents.

Local densities are measured in the Z basis:
\begin{equation}
    \langle n _{i}\rangle = \frac{1}{2}(1 - \langle Z_{i} \rangle).
\end{equation}
The currents on every bond $(i, j)$ can be obtained from density measurements on qubits $i$ and $j$ after performing an additional two-qubit rotation $M_{ij}$ shown in Fig.~\ref{fig:common_circuits}b on every bond within non-overlapping sector. More explicitly, we map $R_Z(-\frac{\phi}{2})_j X_j Y_k R_Z(\frac{\phi}{2})_j \rightarrow \mathds{1}_j Z_k$ and $R_Z(\frac{\phi}{2})_k Y_j X_k R_Z(-\frac{\phi}{2})_k \rightarrow Z_j \mathds{1}_k$ in the current operator from Eq.~\ref{eq:current_op} expressed in terms Pauli matrices.  

After this transformation the expectation value of the current on a single bond can be computed as:
\begin{equation}
    \langle J _{jk}\rangle = -J(\langle \mathds{1}_j Z_{k} \rangle - \langle Z_{j} \mathds{1}_k\rangle).
\end{equation}
To obtain current measurements for the full system we measure $\langle J _{jk}\rangle$ over four circuit realizations, each corresponding to one of the sectors mentioned above.

Because of the stochastic nature of the driving, each circuit corresponds a different trajectory and a different final density and current configuration. In order to compute expectation values, the results of the measurements are averaged over multiple circuit realizations.

\subsection{Fermionic circuit}
\begin{figure}[t]
    \centering
    \includegraphics[width=\columnwidth]{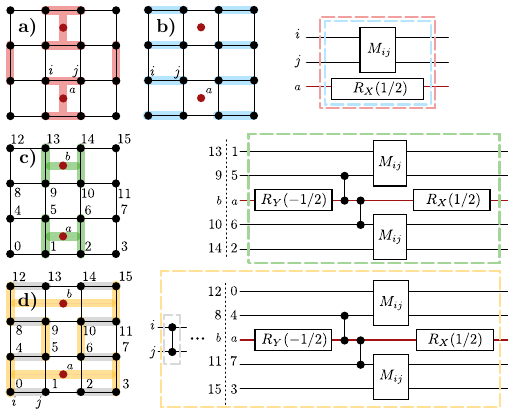}
    \caption{Two-qubit rotations to measure fermionic currents in four different circuit sectors, shown in red, blue, green and yellow. Circuits applied to each sector are shown in dashed boxes of the same colour.  In (a)-(b) coloured pairs of neighbouring qubits and ancillas are labelled $i$, $j$ and $a$ respectively. In sectors (c)-(d) all bonds are numbered with the same convention as Fig.~\ref{fig:hybrid_mapping_detailed}. Note that in sector (d) currents are measured on yellow bonds only, but to apply the required two-qubit rotations on the qubits involved, we need to apply a series of CZ gates on all neighbouring qubits $i$ and $j$ coloured in gray.}
    \label{fig:fermions_2qr}
\end{figure}

\begin{figure*}[t]
    \centering
    \includegraphics[width=\linewidth]{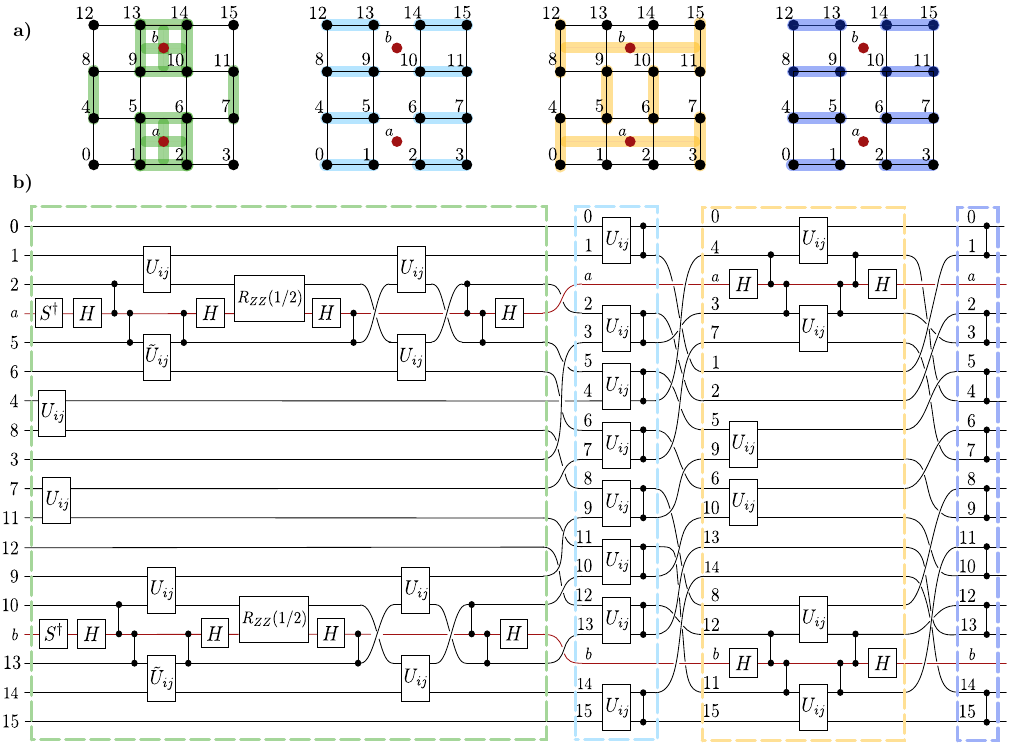}
        \caption{Fermionic Trotter step for each sector of bonds. All sites are numbered from ``0" (source) to ``15" (drain) and the two ancillas are labelled ``a" and ``b", as in Fig.~\ref{fig:hybrid_mapping_detailed}. (a) Different sectors are shown in different colours on the lattice in the same order that they are applied (left to right). (b) shows the full Trotter circuit where each dashed box contains all the gates applied on the same-colour sector from (a). Note that the qubit swaps are not implemented using gates but are implemented by physically moving the ions.}
    \label{fig:fermionic_trotter}
\end{figure*}

To prepare the initial fermionic state, we fix a random density pattern for the physical qubits and in addition, we need to ensure that the stabilizer of the mapping is in the correct $+1$ subspace. This requires rotating the ancillas in the Y basis. In practice, we condition an $R_X$ rotation on the ancillas depending on the state of the adjacent physical qubits belonging to the stabilizer from Eqn.~\ref{eq:stabilizer}:
\begin{figure}[H]
    \centering
    \includegraphics[width=\columnwidth]{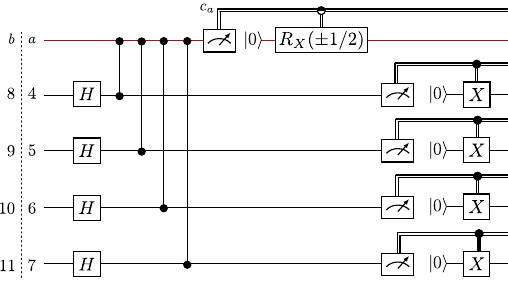}
\end{figure}
All other qubits are prepared following the circuit in Fig.~\ref{fig:common_circuits}a.
Similarly to the bosonic circuit, we subdivide the bonds in commuting sectors on which we can apply gates in parallel. The choice of sectors is made to minimise circuit depth.
To evolve the full system unitarily we implement the mapping above one sector at a time, starting from the bonds close to the ancillas, as shown in Fig.~\ref{fig:fermionic_trotter}a. On the horizontal (vertical) bonds belonging to the green sector, we perform the change of basis $HS^\dagger Z=Y$ ($HZ=X$) on the ancillas, entangle each ancilla to one qubit per neighbouring bond using CZ gates and apply unitary two-qubit gates on each bond. Depending on the relative position of the bonds with respect to the ancilla, the gate will be either $U_{ij}$ or $\tilde{U}_{ij}$, as shown in Fig.~\ref{fig:fermionic_trotter}b. Then, we disentangle the qubits and reverse the rotation of the ancillas. Note that in the circuit we used the fact that $HSH \propto R_X(1/2)$ to combine the evolution of the horizontal and vertical bonds.

In the light blue sector we apply the gate $U_{ij}$ on all bonds, followed by CZ gates. These are necessary to be able to act on the bonds in the yellow sector with the same gate set as the vertical bonds in the green sector.  Note that the qubit swaps in the circuit are performed by physically moving the ions and are not decomposed into gates. The trotter evolution terminates with reverting the CZ operation on the bonds in the dark blue sector.
Each full Trotter evolution is then followed by the driving protocol shown in Fig.~\ref{fig:setup_and_circuit}b.

After $m$ timesteps, we perform full-circuit measurements to extract densities and currents. The density measurements are performed in the computational basis on all qubits simultaneously. To measure currents, we need to perform the two-qubit rotations from Fig.~\ref{fig:common_circuits}b and project again into the correct physical subspace. To achieve this, we group the bonds in non-overlapping sectors and apply the associated gate  sequence shown in Fig.~\ref{fig:fermions_2qr}.

\section{Lindblad derivation}\label{sec:lindblad derivation}
Here we include a brief derivation of Eq.~\eqref{eq:linblad} as the $\Delta t\rightarrow 0$ limit of Eq.~\eqref{eq:CPTP}. We proceed by expanding $\rho(t + \Delta t)$ in powers of $\Delta t$ around $\rho(t)$, to compute the derivative $\text{d}\rho/\text{d}t$. For convenience, we will change $\Delta t \rightarrow dt$, and $n\Delta t \rightarrow t$

First, let us define the driving rate via $p = \gamma dt$. Then, if we absorb the unitary time evolution into our Kraus operators, our channel becomes
\begin{equation}
\rho(t+dt) = \sum_k K_k \rho(t) K^\dag_k,
\end{equation}
where the Kraus operators are
{\allowdisplaybreaks
\begin{align}
    K_0 &= (1-\gamma dt) e^{-iHdt} = \mathds{1} - \gamma dt \mathds{1} - idt H + O(dt^2),\nonumber \\
    K_1 &= \sqrt{\gamma dt} c^\dagger_s + O(dt), \nonumber\\
    K_2 &= \sqrt{\gamma dt} n_s + O(dt), \nonumber\\
    K_3 &= \sqrt{\gamma dt} (1-n_d) + O(dt), \nonumber\\
    K_4 &= \sqrt{\gamma dt} c_d + O(dt), \nonumber\\
    K_5 &= \gamma dt c^\dag_s (1-n_d) +O(dt^2), \nonumber\\
    K_6 &= \gamma dt n_s (1-n_d) + O(dt^2), \nonumber\\
    K_7 &= \gamma dt c^\dag_s c_d + O(dt^2), \nonumber\\
    K_8 &= \gamma dt n_s c_d + O(dt^2). \label{eq:kraus_operators}
\end{align}}  

Since the operators appear quadratically in the channel, the leading contribution for $K_5$ through $K_8$ is at order $dt^2$, which vanishes in the limit for the derivative. $K_0$ on the other hand, due to the presence of the leading identity, contributes non-vanishing terms at order $dt$. Collecting all the relevant terms, we get
\begin{equation}
\begin{aligned}
    \rho(t+dt) = \rho + dt\Big [&-2\gamma \rho -i [H,\rho] +\gamma c^\dag_s \rho c_s + \gamma n_s \rho n_s\\  &+ \gamma (1-n_d) \rho (1-n_d) + c_d \rho c^\dag_d  \Big]\\
    &+O(dt^2).
\end{aligned}
\end{equation}
To get this in the desired form, note that $2\gamma \rho = \{ \gamma \mathds{1}, \rho\}$, to get 
\begin{equation}
    \frac{\text{d}\rho}{\text{d}t} = -i [H,\rho] - \{\gamma\mathds{1}, \rho \} + \sum_i L_i \rho L_i^\dag,
\end{equation}
where we have defined
\begin{equation}
\begin{aligned}
    L_1 &= \sqrt{\gamma} c_s^\dag, \\
    L_2 &= \sqrt{\gamma} n_s, \\
    L_3 &= \sqrt{\gamma} (1-n_d), \\
    L_4 &= \sqrt{\gamma} c_d
\end{aligned}
\end{equation}
Finally, we note that the normalization factor
\begin{equation}
    \gamma \mathds{1} = \frac{1}{2}\sum_i L_i^\dag L_i,
\end{equation}
and hence we can write in the standard Lindblad form
\begin{equation}
\frac{d \rho}{dt} = -i \left[H, \rho \right] + \sum_i \left( L_i \rho L_i^\dagger -\frac{1}{2}\left\{L^\dagger_i L_i, \rho  \right\} \right).
\end{equation}
as in Eq.~\eqref{eq:linblad}.

\section{Optimal driving rate}\label{sec:optimal gamma}
\begin{figure}[t]
    \centering
    \includegraphics[width=\columnwidth]{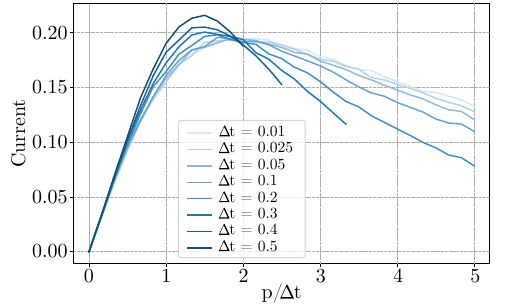}
    \caption{Finding the optimal rate $p/\Delta t$ for the mid-circuit measurement. Sweeps over $p$ values for different values of $\Delta t$ for free-fermions in a $4 \times 4$ system. Optimal rate seems to be $p/\Delta t = 2$, which is shifted lower for larger values of $\Delta t$. We clearly see the Zeno effect when the measurement rate is too high. Current is extracted from the statistics of the Kraus operators.}
    \label{fig:current_sweep}
\end{figure}

In order to get sufficiently large signals to measure, particularly for currents, we wanted to choose the driving rate that maximised the current through the system. For low driving rate $\gamma$, we clearly get low currents due to slow rate of particle injection / removal. However, for large $\gamma$ we also limit currents due to the quantum Zeno effect~\cite{Misra1977} from measuring the corners too frequently. We therefore expect an optimal $\gamma$ that is comparable with the hopping rate $J$. To get a better estimation, we considered a free-fermion simulation on the 4 $\times$ 4 system for different values of $\Delta t$ and measured the net current through the system as a function of $\gamma = p / \Delta t$, shown in Fig.~\ref{fig:current_sweep}. This shows that in the limit $\Delta t \rightarrow 0$ the maximum current is achieved when $\gamma = 2$. For larger values of $\Delta t$, the optimal $\gamma$ is slightly lower but remains close to $\gamma \approx 2$. Furthermore, the value of the current at $\gamma=2$ is approximately independent of $\Delta t$ and so we chose $\gamma = 2$ for all experiments.

\section{Determining experimental parameters}\label{sec:optimal parameters}
\begin{table}[t]
\renewcommand{\thetable}{T1}
    \centering
    \begin{tabular*}{\columnwidth}{@{\extracolsep{\fill}}cccc}
    \toprule
    Setup & $\Delta t$ & $m$ & Trajectories\\
    \midrule
    \parbox[c][2.5em]{10em}{Fermions \linebreak $V=1$, $\phi = \pi/2$}   & 0.29  & 18  & 1480 \\
    \parbox[c][2.5em]{10em}{Fermions \linebreak $V=0$, $\phi = \pi/2$}   & 0.27  & 16  & 1480 \\
    \parbox[c][2.5em]{10em}{HCB \linebreak $V=1.5$, $\phi = 0$}          & 0.31  & 14  & 1280 \\
    \parbox[c][2.5em]{10em}{Fermions \linebreak $V=0$, $\phi = 0$}       & 0.21  & 14  & 1480 \\
    \parbox[c][2.5em]{10em}{HCB \linebreak $V=0$, $\phi = 0$}            & 0.31  & 10  & 1280 \\
    \bottomrule
    \end{tabular*}
    \caption{Final choices of time increment $\Delta t$, timesteps $m$  and trajectories for experimental simulations on the H1-1 device for all setup choices.}
    \label{table:parameter_choices}
\end{table}

\begin{figure*}[p]
    \centering
    \includegraphics[width=0.93\linewidth]{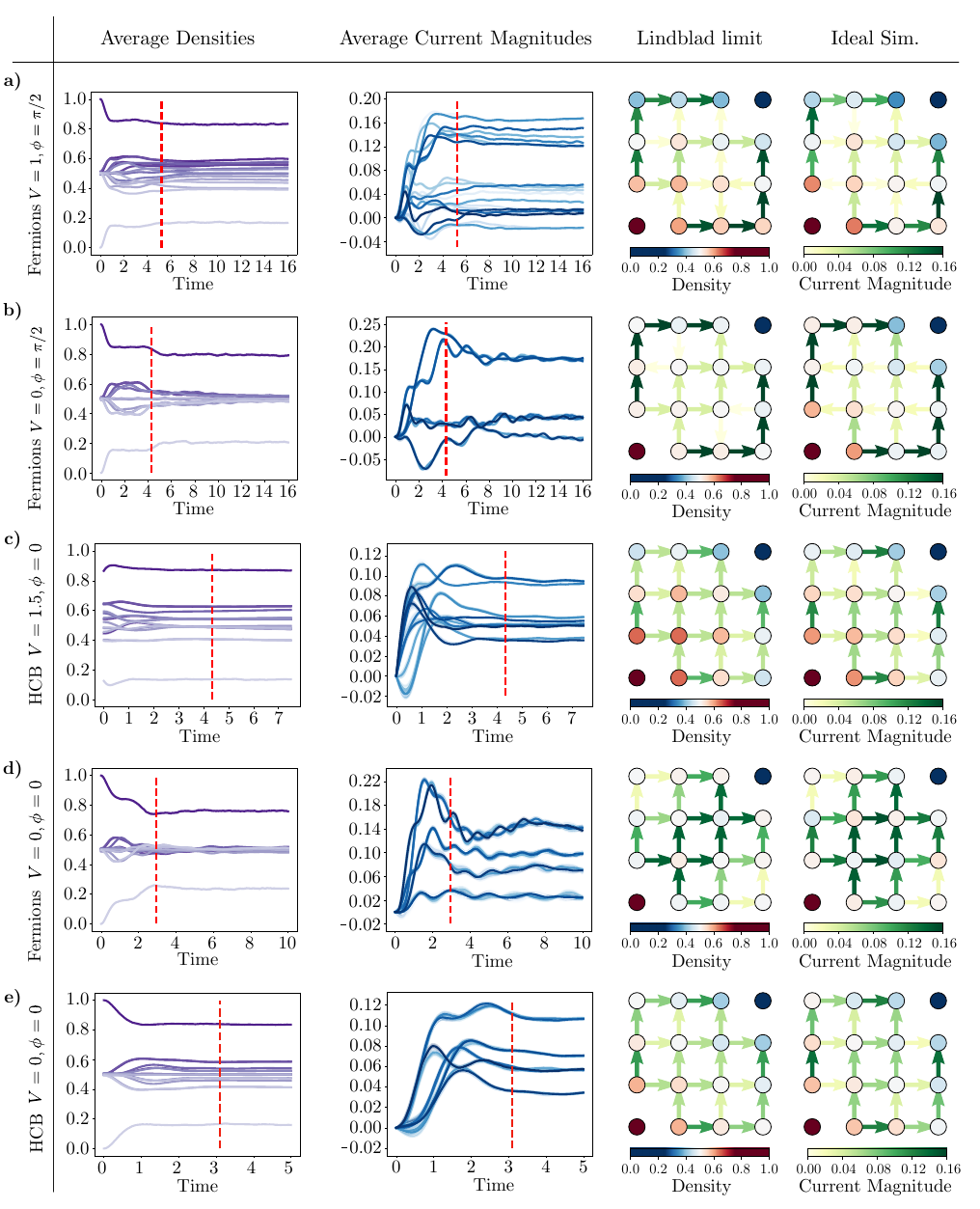}
    \caption{Numerical simulation to determine experimental parameters. In the first two columns from the left, the time evolution close to the Lindblad limit of both average density and average current magnitude. In the density traces, source and drain are shown in dark and light purple respectively. In the experiment we evolve up to a final time $T = m\Delta t$ shown with a dashed red vertical line. In both plots, the simulation parameters are (a)-(b) $m=1000$ and $dt=0.02$, (c) $m=600$ and $dt=0.05$, (d) $m=1000$ and $dt=0.01$, (e) $m=700$ and $dt=0.01$.  The second two columns show the NESS snapshots at arbitrary long times $t=mdt$ in the same Lindblad limit and the snaphots close to the NESS at time $T$ from the ideal simulation with the chosen values of $m$ and $\Delta t$ as the experiment. All currents and densities are averaged over 10000 trajectories.} 
    \label{fig:parameter_sweep}
\end{figure*}

The experimental choice of parameters is based on numerical simulations. For each choice of interaction strength $V$ and magnetic flux $\phi$, the goal is to choose a time increment $\Delta t$ that strikes a balance between low Trotter error, long enough run times to reach the NESS and shallow circuits to reduce device error. 
To do so, we evolved close to the Lindblad limit for arbitrarily long times, and select the smallest time $T$ when both average density and average currents have settled. This indicates that the system has approximately reached the steady state.
After selecting the minimum evolution time $T$, we sweep through values of $\Delta t$ and number of timesteps $m \approx T / \Delta t$ and choose the largest $\Delta t$ such that the trotter simulation qualitatively reproduces the Lindblad limit.
\\The final choices of $m$ and $\Delta t $ for all different setups are shown in Table ~\ref{table:parameter_choices}.

Fig.~\ref{fig:parameter_sweep} shows the density and current traces in the Lindblad limit for all the experimental cases executed on Quantinuum's H1-1 device, together with the NESS both in the Lindblad limit and in the ideal simulation with the chosen experimental parameters. The former evolves unitarily under the full system Hamiltonian from Eq. ~\ref{eq: Hamiltonian}, while the latter follows from the Trotter circuit. In both simulations, the non-unitary drive is implemented by applying Kraus operators from ~\ref{eq:kraus_operators} on source and drain. Average densities and currents are then calculated from expectation values for each trajectory.

Both bosonic densities and currents (Fig.~\ref{fig:parameter_sweep}c-e) settle much faster than their fermionic counterparts, allowing us to select larger values of $\Delta t $ while keeping high fidelity to the physics in the continuous time limit.

In fermionic circuits, the currents are more localised and take a longer time to resolve. In these cases, we had to make some compromises on the choice of final time $T$ (shown with dashed red lines) and time increment $\Delta t$: we chose to run for long enough times to move past the large initial current peaks and still observe NESS features, but we stopped before some of the densities and currents had the time to fully settle. This allowed us to keep $\Delta t$ small enough to limit the size of the Trotter error.

Even if the currents haven't fully stabilised, Fig.~\ref{fig:parameter_sweep}b and Fig.~\ref{fig:parameter_sweep}c both display clear signatures of the ballistic current transport in fermionic systems.

\section{Interacting SSEP Simulation}\label{sec:SSEP}

\begin{figure}[t]
    \centering
    \includegraphics[width=1\columnwidth]{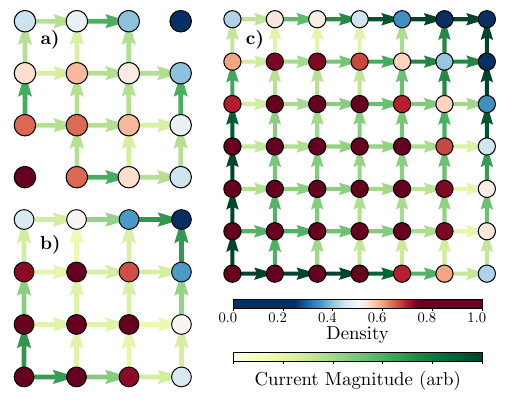}
    \caption{SSEP  and HCB simulations from classical numerics. (a) shows the NESS for HCB with $V=1.5$ in the Lindblad limit, obtained with $dt=0.05$, $m = 600$ timesteps and averaged over 10000 trajectories. (b) and (c) show the NESS for classical stochastic interacting SSEP ($V=-1$) in a $4 \times 4$ lattice and $7 \times 7$ lattice respectively evolved for $m=300$ timesteps and averaged over $10^6$ trajectories.} 
    \label{fig:SSEP}
\end{figure}

\begin{figure}[t]
    \centering
    \includegraphics[width=1\columnwidth]{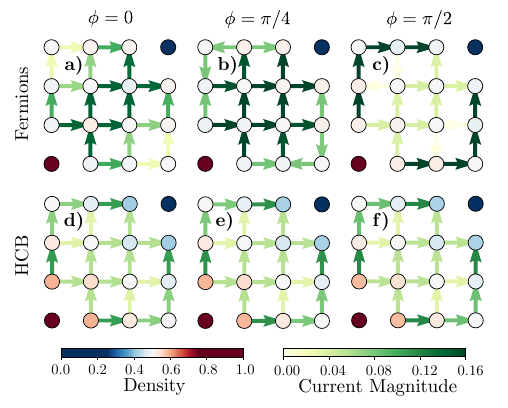}
    \caption{Numerical results for final snapshots of the NESS for HCB and fermions with $V=0$ and different values of magnetic flux $\phi$. For all HCB snapshots: $m=700$ timesteps, $dt=0.01$ and $10000$ trajectories. For the fermionic snapshots: (a)-(b) $m=1000$ timesteps, $dt=0.01$ and $10000$ trajectories, (c) $m=1000$ timesteps, $dt=0.02$ and $10000$ trajectories.} 
    \label{fig:Magnetic_Field}
\end{figure}

To back up our energetic argument for the observed density pattern in Fig.~\ref{fig:hydro}b when introducing interactions, we performed a classical stochastic simulation of an interacting SSEP, for which we provide details here.

The Symmetric Simple Exclusion Process is a classical stochastic model for the local dynamics of particles on a lattice subject to a hard-core constraint (no double occupancy). Here we introduce a discrete time implementation of this model that additionally includes particle injection / removal at the corners and an energetic interactions term.

The basic SSEP is a purely dynamical model with the following local rules
\begin{equation}
    \circ \circ \overset{p}{\rightharpoonup} \circ \circ, \qquad
    \bullet \circ \underset{p}{\overset{p}{\rightleftharpoons}} \circ \bullet, \qquad
    \bullet\bullet \overset{p}{\rightharpoonup} \bullet \bullet.
\end{equation}
The middle rule corresponds hopping of particles, and the probability $p$ of the hop taking place is symmetric. In our implementation we use discrete times. At each time we choose a bond on the lattice uniformly at random, and then we perform the allowed move (effectively swapping the two sites) with probability $p=1$. We define a unit of time to be $N$ of these discrete time steps, where $N$ is the number of sites.

To implement the injection and removal of particles, we add new process at the source and drain sites in the opposite corners. At each time step, with probability $\gamma/ (M+ 2\gamma)$, where $M$ is the number of bonds, we select the source site and set it to be filled, regardless of the current state. Similarly, with the same probability, we select the drain site and set it to be empty. Otherwise, we select a random bond uniformly at random as before (hence with probability $1/(M+2\gamma)$ for each bond).

Finally, to add interactions to this stochastic model, we add an energy term $-V$ for each bond in the configuration $\bullet \bullet$. We then accept or reject any new configuration at each time step using a Metropolis-Hastings acceptance rate
\begin{equation}
    A = \min \left(1, e^{E_\text{old} -E_\text{new}} \right).
\end{equation}
Note that if we inject or remove a particle we always accept the new configuration.

In Fig.~\ref{fig:SSEP} we compare the observed NESS for HCB in the presence of interactions with our classical stochastic interacting SSEP. We see that we recover a similar density pattern, namely with the system being filled more on average except along the top and right edges that are connected to the drain. We observe similar behaviour for both the $4\times 4$ system, as well as a large $7 \times 7$ simulation of the SSEP. Since all coherent effects from the quantum model are absent in our SSEP, we conclude that the density pattern is the result of the energetic cost of configurations. The lower coordination number along the edges compared with the bulk results in a lower energetic barrier for holes along those two edges. 

\AtEndDocument{\renewcommand{\thepage}{\roman{page}}}

\section{Magnetic fields}\label{sec:HCB field}

\begin{figure}
    \centering
    \includegraphics[width=\columnwidth]{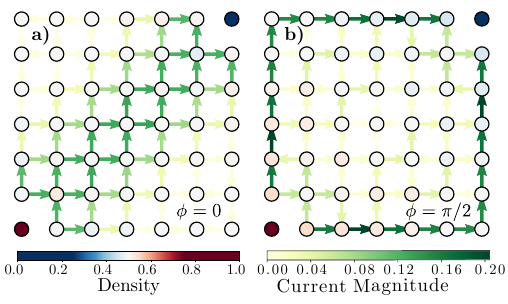}
    \caption{Final time snapshots of average density and instantaneous currents for free-fermions ($V=0$) in a $7 \times 7$ square lattice. (a) $\phi=0$ and (b) $\phi=\pi/2$ are obtained with $dt=0.02$, $m=2000$ timesteps, averaged over 10000 trajectories.}  \label{fig:FF_size_up}
\end{figure}

While non-interacting fermions are very sensitive to magnetic flux strength, HCB display no significant changes in both average density and currents in the NESS. 

Fig.~\ref{fig:Magnetic_Field} shows the final time snapshots of the NESS for both statistics with $\phi = 0$, $\phi=\pi/2$ and the intermediate value $\phi=\pi/4$. As we increase the field strength, the fermionic currents along the diagonal broaden, and eventually they are pushed to the system edges. Bosonic currents display no significant changes in both magnitude and direction of flow for all three values of flux.

This strong dependence on magnetic field strength for fermionic statistics persists in bigger system sizes, confirming that this behaviour is physical and not a simple finite-size effect. As an example, Fig.~\ref{fig:FF_size_up} shows the final time snapshots of the steady state in a $7\times 7 $ square lattice in the free-fermion limit. For both $\phi=0$ and $\phi=\pi/2$ we observe the expected out-of-equilibrium behaviour resulting from ballistic transport.

\putbib
\end{bibunit}

\end{document}